\documentclass[twocolumn,showpacs,preprintnumbers,amsmath,amssymb]{revtex4}

\usepackage{graphicx}
\usepackage{dcolumn}
\usepackage{bm}

\begin{document}


\title{Charm Quark Energy Loss in QCD Matter}
\author{W.C. Xiang$^{1}$, H.T. Ding$^{1}$, D.C. Zhou$^{1,2}$, D. Rohrich$^{2}$ \\
$^1$Institute of Particle Physics, Huazhong Normal University,
Wuhan P.R. China 430079\\
$^2$Department of Physics, University of Bergen, Norway\\ }

\date{\today}

\begin{abstract}
{\em \noindent The energy loss of heavy quarks in a quark-gluon
plasma of finite size is studied within the light-cone path
integral approach. A simple analytical formulation of the
radiative energy loss of heavy quarks is derived. This provides a
convenient way to quantitatively estimate the quark energy loss.
Our results show that if the energy of a heavy quark is much
larger than its mass, the radiative energy loss approaches the
radiative energy loss of light quarks.}
\end{abstract}

\pacs{25.75.-q,12.38.Mh,14.65.Dw}
\maketitle


In high energy heavy ion collisions hard scattering of partons
occur in the early stages of the reaction, well before a
quark-gluon plasma(QGP) might have been formed, producing fast
partons that propagate through the hot and dense medium and lose
their energy. Hard hadronic probes have long been thought to
detect the formation of a quark-gluon plasma in ultrarelativistic
 heavy ion collision. Heavy quark radiative energy loss is such a probe to study the
properties of the quark-gluon plasma. In recent years, the
investigation of the parton energy loss in QCD matter has created
considerable interest$^{[1-7]}$. The study of the induced gluon
radiation from heavy quarks in QGP matter is of great importance
for the understanding of experiment data from high energy
nucleus-nucleus (AA) collisions at the Relativistic Heavy Ion
Collider (RHIC) and the Large Hadron Collider (LHC). Recent
measurements$^{[8-13]}$ of high $p_\bot$ hadron production and its
centrality dependence in Au-Au collisions at
$\sqrt{s_{NN}}=200~$GeV provide the first evidence for
medium-induced parton energy loss. The motivation of this paper is
to employ light cone path integral (LCPI) approach to solve the
massive partons radiative energy loss. The analytical solution is
obtained by the way of light cone path integral. Our analytical
result is consistent with the former numerical result-the parton
energy loss
tends to be smaller for massive quarks than for massless ones.\\

\indent In 1953 Landau and Pomeranchuk$^{[14]}$ predicted with
classical electrodynamics if the formation length of the
bremsstrahlung becomes comparable to the distance over which the
multiple scattering becomes important, the bremsstrahlung will be
suppressed. Migdal$^{[15]}$ developed a quantitative theory of
this phenomenon. In current literature, we call the suppression of
radiation processes in medium the Landau-Pomeranchuk-Migdal (LPM)
effect. First results on the LPM effect in QCD were given by
Gyulassy and wang$^{[16,17]}$, they first discussed that the
parton jet, produced in inelastic collision, propagating in the
QCD matter will lose its energy due to medium-induce gluon
radiation(G-W model). They pointed out, comparing with elastic
energy loss, the contribution of inelastic energy loss is more
important. G-W model has been extended by R.Baier, Y.Dokshitzer,
A.H.Mueller, S.Peigne and D.Schiff(BDMPS), using equal-time
perturbation theory. The calculation of BDMPS about the energy
loss of inelastic scattering indicate that the parton radiative
energy loss has a square dependence on the path length in the
medium$^{[18]}$. Based on G-W model, M.Gyulassy, P.Levai and
I.Vitev(GLV) developed opacity technology to calculate jet energy
loss$^{[19]}$. E.K.Wang and X.N.Wang first discussed the detailed
balance effect of jet energy loss in hot QGP medium$^{[3]}$. They
pointed out the modified energy dependence of the energy loss will
affect the suppression shape of moderately high $p_t$ hadrons due
to jet quenching in high energy heavy ion collision. All above
mentioned approaches are focused on massless parton energy loss.
Based on the non-abelian Furry approximation, U.A.Wiedemann
obtained a medium-induce gluon distribution$^{[20]}$. It is a
general result and provides a proof of the equivalence of the
BDMPS and Zakharov formalisms. They extended medium-induce gluon
distribution to calculate quenching weights$^{[21]}$ and massive
quark energy loss$^{[6]}$ and got
significative information for jet quenching.\\

\indent Due to the mass effect, it is hard to solve the problem of
massive parton emission. Up to now, most results of heavy quark
energy loss have been numerical. The light-cone path integral
approach is a simple way of dealing with bremsstrahlung of photons
and gluons and it can give an analytical result. In the
path-integral formalism, the radiation cross section is determined
by a dipole cross section which essentially measures the
difference between elastic scattering amplitudes of different
projectile Fock state components as a function of impact
parameter. Baier, Dokshitzer and Schiff (BDMS) have shown$^{[22]}$
that the evolution of the rescattering amplitude in the
BDMS-formalism is determined by Zakharov's dipole cross
section$^{[20]}$. Accordingly, static Debye screened scattering
centers are considered and all the scattering centers are supposed
to be independent. The probability of gluon emission in the LCPI
approach is expressed through the solution of a two-dimensional
Schr\"odinger equation with an imaginary potential. The
two-dimensional Hamiltonian reads$^{[23]}$:
\begin{equation}
H=-\frac{1}{2M(x)}(\frac{\partial}{\partial{\boldsymbol{\rho}}})^2-i\frac{n(z)\sigma_3(\rho,x)}{2},
\end{equation}
where $M(x)=Ex(1-x)$, $x$ is the gluon fractional momentum, $n(z)$
is the number density of the medium at the longitudinal coordinate
z and $\sigma_3$ is the cross section of the interaction of a
color singlet
$q{\bar q}g$ system with a color center.\\

\indent The contribution of the bremsstrahlung mechanism to the
cross section of gluon production can be written as$^{[7]}$:
\begin{equation}
\frac{d\sigma_{eff}^{BH}(x,z)}{dx}=Re\int
d{\boldsymbol{\rho}}\psi^{\ast}({\boldsymbol{\rho}},x)\sigma_3({\rho},x)\psi({\boldsymbol{\rho}},x,z),
\end{equation}
where $\psi({\bf\rho},x)$ is the light-cone wave function for the
$q\to qg$ transition in vacuum and $\psi({\bf\rho},x,z)$ is the
medium-modified light-cone wave function for the $q\to qg$
transition in medium at the longitudinal coordinate z,
${\boldsymbol{\rho}}$ is the transverse coordinate and $x$ is the
Feynman variable of the radiated gluon. The wave functions
reads$^{[7]}$:
\begin{eqnarray}
\psi({\boldsymbol{\rho}},x)&=&p(x)(\frac{\partial}{\partial
\rho_x^\prime}-is_g\frac{\partial}{\partial
\rho_y^\prime})\nonumber\\
&&\times\int_0^\infty d\xi\exp(-\frac{i\xi}{L_f})
K_0({\boldsymbol{\rho}},\xi|{\boldsymbol{\rho}^\prime,0})|_{{\boldsymbol{\rho}^\prime}=0};
\end{eqnarray}
\begin{eqnarray}
\psi({\boldsymbol{\rho}},x,z)&=&p(x)(\frac{\partial}{\partial
\rho_x^\prime}-is_g\frac{\partial}{\partial
\rho_y^\prime})\nonumber\\
&&\times\int_0^z d\xi\exp(-\frac{i\xi}{L_f})
K_0({\boldsymbol{\rho}},z|{\boldsymbol{
\rho}^\prime},z-\xi)|_{{\boldsymbol{\rho}^\prime}=0}\nonumber,\\
\end{eqnarray}
where $p(x)=i\sqrt{\alpha_s/2x}~[s_g (2-x)+2s_q x]/2M(x)$,
 $s_{qg}$ denotes parton helicities ($s_q=\frac{1}{2}$, $s_g=\pm1$), $K_0$ is the Green function for
 the two-dimensional Hamiltonian. $K_0$ can be written as:
\begin{equation}
K_0({\boldsymbol{\rho}_2},z_2|{\boldsymbol{\rho}_1},z_1)=\frac{M(x)}{2\pi
i(z_2-z_1)}\exp[\frac{iM(x)({\boldsymbol{\rho}_2}-{\boldsymbol{\rho}_1})^2}{2(z_2-z_1)}],
\end{equation}
where $L_f=2Ex(1-x)/\epsilon^2$,$\epsilon^2=m_g^2 (1-x)+m_q^2
x^2$, $L_f$ is the gluon formation length.
 $m_q$ is the quark mass and $m_g$ is the mass of the radiated gluon.
 The latter plays the role of an infrared cut-off removing
 contributions of the long-wave gluon excitations which cannot be
 treated perturbatively. We assume that the heavy quark is produced in the
 central rapidity region at $\eta=0$ and the production point is
 at $z=0$, propagating through a hard mechanism in a medium of extent $L$ along
 the $z$ axis. The induced gluon bremsstrahlung spectrum can be
 represented as:
 \begin{equation}
 \frac{dp}{dx}=\int_0^L dzn(z)\frac{d\sigma_{eff}^{BH}(x,z)}{dx},
 \end{equation}
 where $n(z)$ is the number density of the medium.\\

\indent Using the formulas (3)-(5), it is easy to obtain the
light-cone wave function after a simple
 calculation:
 \begin{eqnarray}
 \psi(\rho,x)&=&\frac{p(x)M^2(x)}{2\pi}(-\rho_x+is_g\rho_y)\nonumber\\
 &&\times\int_0^\infty\frac{d\xi}{\xi^2}\exp(-\frac{i\xi}{L_f})
 \exp[\frac{iM(x)\boldsymbol{\rho}^2}{2\xi}];
 \end{eqnarray}
 \begin{eqnarray}
 \psi(\rho,x,z)&=&\frac{p(x)M^2(x)}{2\pi}(-\rho_x+is_g\rho_y)\nonumber\\
 &&\times\int_0^z\frac{d\xi}{\xi^2}\exp(-\frac{i\xi}{L_f})
 \exp[\frac{iM(x)\boldsymbol{\rho}^2}{2\xi}].
 \end{eqnarray}
Substituting (7) and (8) to (2), the cross section of gluon
production can be represented as:
\begin{eqnarray}
\frac{d\sigma_{eff}^{BH}(x,z)}{dx}&=&\frac{p^2(x)M^3(x)}{2\pi^2}Re\int
d{\bf\rho}\rho\epsilon
K_1(\rho\epsilon)i\sigma_3(\rho,x)\nonumber\\
&&\times\int_0^z\frac{d\xi}{\xi^2}\exp(-\frac{i\xi}{L_f})
\exp[\frac{iM(x)\boldsymbol{\rho}^2}{2\xi}],
\end{eqnarray}
with $K_1(\rho\epsilon)$ the modified Bessel function of the
second kind, $\sigma_3(\rho,x)$ is the three-body cross section of
the imaginary potential$^{[24]}$,
$\sigma_3(\rho,x)=C_A/2C_F[\sigma_2((1-x)\rho)+\sigma_2(\rho)-\frac{1}{C_A^2}\sigma_2(x\rho)]=C_3(x)\rho^2$,
where $\sigma_2(\rho)$ is the dipole cross section for scattering
of a $q\overline q$ pair on a color center, $C_3(x)=C_2(\rho)A(x)$
with $A(x)=[1+(1-x)^2-x^2/N_c^2]C_A/2C_F$. In the region
$\rho<<\frac{1}{\mu}$ (here $\mu$ is the Debye Screening mass),
which dominates the spectrum for strong suppression, $C_2(\rho)$
takes the form$^{[7]}$ :
\begin{equation}
C_2(\rho)\approx\frac{C_FC_T\alpha_s^2\pi}{2}\ln(\frac{1}{\rho\mu}),
\end{equation}
where $C_A$, $C_F$, $C_T$ are the color Casimir operators. From
(10), we can
see that $C_2(\rho)$ has a slow logarithmic dependence on $\rho$.\\

\indent In order to derive a quantitative estimate, we take the
charm quark of mass $m_q=1.5~$GeV. When $\rho\epsilon$ is small,
we can expand the Bessel function $K_1(\rho\epsilon)$ and keep the
first two terms. After a complex algebra calculation, one can
obtain the main contribution of the bremsstrahlung to the cross
section of gluon production:
\begin{eqnarray}
\frac{d\sigma_{eff}^{BH}(x,z)}{dx}&=&\frac{\alpha_s^2
C_FC_TA(x)G(x)}{8M(x)}\nonumber\\
&&\times\left\{\frac{\pi}{2}L_f\sin\frac{z}{L_f}+L_f(1-c)(\cos\frac{z}{L_f}-1)\right.\nonumber\\
&&+\left.L_f\ln\frac{M(x)}{2\mu^2z}
(1-\cos\frac{z}{L_f})+\frac{z^2}{4L_f}\right\},
\end{eqnarray}
where $G(x)=\alpha_sC_F[1-x+x^2/2]/x$ and the Euler constant
$c\simeq0.5772$. The radiative energy loss can be written as:
\begin{equation}
\Delta E=\int_{\omega_{cr}}^E d\omega\omega\frac{dp}{d\omega}.
\end{equation}
Considering light partons, the mass can be neglected,
$L_f\to\infty$, $\cos\frac{\xi}{L_f}\approx1$,
 $\sin\frac{\xi}{L_f}\approx0$. From (11) and (12) we obtain the light quark
radiative energy loss $\Delta E$ $^{[7]}$:
\begin{equation}
\Delta
E=\frac{C_F\alpha_s}{4}\frac{L^2\mu^2}{\lambda_g}\ln\frac{E}{\omega_{cr}},
\end{equation}
where $\frac{1}{\lambda_g}=\alpha_s^2\pi C_FC_TA(0)n/2\mu^2$. At
$E\to\infty$ the energy loss (13) is equal to GLV's
result$^{[19]}$:
\begin{equation}
\Delta
E_{GLV}=\frac{C_F\alpha_s}{4}\frac{L^2\mu^2}{\lambda_g}\ln\frac{E}{\mu},
\end{equation}
where $\mu$ is the Debye Screening mass and
$\omega_{cr}\sim\max(nC_3L^3/4,L\mu^2/2)$. Formula (14) reflect
the logarithm energy dependence of radiative energy loss.\\

\indent Now let us discuss the radiative energy loss of a heavy
quark with formula (11). Considering the quark mass effect, the
gluon formation length $L_f$ is a finite quantity. In formulas
(3), (4), (7), (8) and (9), the $\exp(-\frac{i\xi}{L_f})$ is no
longer equal to $1$ as in the case of a light parton. The
mass-dependence of the gluon distribution (6) comes from the phase
factor $\exp(-\frac{i\xi}{L_f})$, which is analogous to the mass
dependence of the gluon distribution which comes from the phase
factor $\exp[i\overline{q}(y_l-\overline{y_l})]$,
$\overline{q}=x^2m_q^2/2\omega$$^{[3]}$. In the high energy limit,
the formation length, $L_f$, becomes larger than the quark path
length in the QGP, i.e.$L_f>>L$$^{[16]}$. Using (6), (11) and
(12), we can get:
\begin{eqnarray}
\Delta
E&=&\frac{C_F\alpha_s}{4}\frac{L^2\mu^2}{\lambda_g}\left\{\ln\frac{E}{\omega_{cr}}\right.\nonumber\\
&&+\frac{m_g^2L}{3\pi\omega_{cr}}(1-\frac{\omega_{cr}}{E}\ln\frac{E^2}{2\mu^2L\omega_{cr}}
+\ln\frac{\omega_{cr}}{2\mu^2L})\nonumber\\
&&+\left.\frac{m_q^2L}{3\pi
E}(-\frac{\pi^2}{6}-\frac{\omega_{cr}}{E}\ln\frac{\omega_{cr}}{2\mu^2L}+\ln\frac{E}{2\mu^2L})\right\},\nonumber\\
\end{eqnarray}
where $m_g=0.375~$GeV is the mass of the related gluon. From (15)
we can see that the first term is the radiative energy loss of a
light quark. Considering the quark mass effect, the latter terms
are the modifications of the light quark.
\begin{figure}
\includegraphics[totalheight=2.3in]{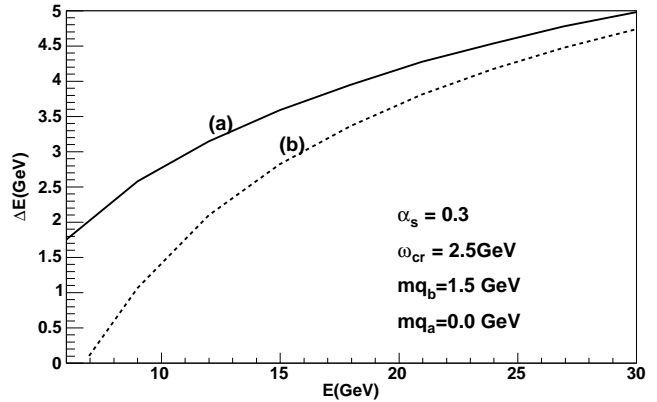}
\centering \caption{The radiative energy loss of light quarks and
heavy quarks as a function of their energy in a plasma
characterized by $\mu=0.5~$GeV, $\lambda_g=1~fm$, $m_g=0.375~$GeV
and $L=4~fm$. The solid line $(a)$ is for light quarks, which were
studied by Zakharov[7]; the dashed line $(b)$ is for heavy quarks
studied by us.}
\end{figure}

\begin{figure}
\includegraphics[totalheight=2.3in]{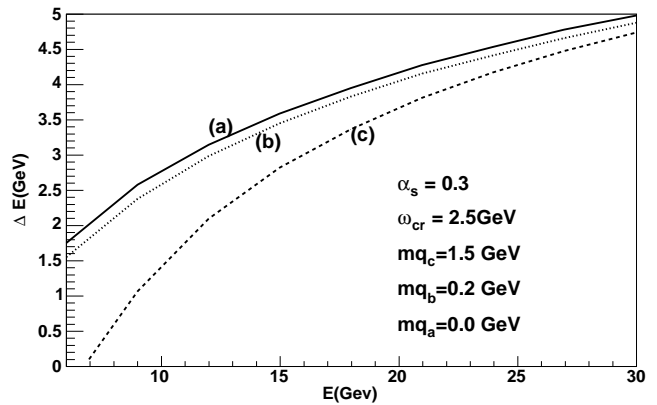}
\centering \caption{The radiative energy loss of light quarks and
heavy quarks as function of their energy in a plasma characterized
by $\mu=0.5~$GeV, $\lambda_g=1.0~fm$, $m_g=0.375~$GeV and
$L=4~fm$. The solid line $(a)$ is for light quarks, which were
studied by Zakharov[7]; the dotted line $(b)$ for a quark of
$m_q=0.2~$GeV; the dashed line $(c)$ is the energy loss for charm
quarks via Eq.(15). It is shown $\Delta E$ is dependent on the
quark mass, the radiative energy loss tends to be smaller for
heavy quarks than for light quarks.}
\end{figure}

\indent For the sake of seeing the suppression of the heavy quark
energy loss more clearly, we use Fig.1 and Fig.2 to show the
radiative energy loss of light quarks and heavy quarks in a QGP
and the dependence on the mass. Fig.1 shows the radiative energy
loss of light quarks and heavy quarks. In Fig.1 we can see that
the difference between light quarks and heavy quarks becomes
smaller with increasing energy of the quark. It is obvious that
when the quark energy is much larger than its mass, the heavy
quark energy loss is approaching the one of light quarks. At high
energies, our results are consistent with Gyulassy's calculations
$^{[25]}$. To illustrate the quark mass dependence of $\Delta E$,
we use Fig.2 to compare the results for a heavy quark (charm quark
$m_q=1.5~$GeV) to light quarks with mass $m_q=0$ and
$m_q=0.2~$GeV. With the above values of $m_q$, one can see the
extent of the $\Delta E$ dependence on the quark mass. Both Fig.1
and Fig.2 show that the formulas (13),(14) and (15) are applicable
in the high energy limit. At RHIC ($\sqrt{s_{NN}}=200~$GeV) and
LHC ($\sqrt{s_{NN}}=5.5~$TeV) energies, all the above formulas are
operable. \\

\indent In this work, we assume static Debye screened scattering
centers and that all the centers are independent. The energy of
the parton is supposed to be high enough so that the condition
$L_f>>L$ is fulfilled. We use the light-cone path integral method
to deal with the gluon emission and thus obtain a simple
analytical formula for the heavy quark radiative energy loss.
Equation $(15)$ can be used to estimate the measurable yield of
hadrons containing heavy quarks. To our knowledge, up to now most
of the calculations of the heavy quark radiative energy loss have
been carried out by numerical simulations. In the high energy
limit our results are consistent with Gyulassy's numerical
calculations$^{[25]}$. But one should keep in mind that when the
quark mass becomes larger, $\rho\epsilon$ is no longer a small
quantity and the Bessel function $K_1(\rho\epsilon)$ cannot be
expanded as above. In a following work, we will try to extend our
result to accommodate all quark masses.

\indent We gratefully acknowledge Prof. B.G. Zakharov, Dr. B.W.
Zhang, T. Wu and F.L. Liu for their very helpful discussions. This
work was supported partly by the fund of National Natural Science
Foundation of China under the grant No. 10075022, partly by the
fund of National Education Ministry of China and in part by the
Guest Researcher Programme, University of Bergen, Norway.

\newpage 

\end{document}